\documentclass[11pt, a4paper]{article}   	

\pdfoutput=4 

\usepackage{jcappub} 
\usepackage{graphicx}
\usepackage{amssymb}
\usepackage{epstopdf,braket}
\usepackage{amsmath,mathtools}
\usepackage{multirow}
\usepackage{siunitx,booktabs}
\usepackage{comment,xcolor}
\usepackage{hyperref}
\newcommand{\carlo}[1]{{\leavevmode\color{black}#1}}
\newcommand{\andreafin}[1]{{\leavevmode\color{black}#1}}

\newcommand{\andreamod}[1]{{\leavevmode\color{black}#1}}

\AtBeginDocument{
\heavyrulewidth=.08em
\lightrulewidth=.05em
\cmidrulewidth=.03em
\belowrulesep=.65ex
\belowbottomsep=0pt
\aboverulesep=.4ex
\abovetopsep=0pt
\cmidrulesep=\doublerulesep
\cmidrulekern=.5em
\defaultaddspace=.5em
}
\begin{document}

\title{The importance of observing astrophysical tau neutrinos}

\title{The importance of observing astrophysical tau neutrinos}

\author[a]{Andrea Palladino}
\affiliation[a]{Deutsches Elektronen-Synchrotron (DESY), Platanenallee 6, D-15738 Zeuthen, Germany}
\author[b]{Carlo Mascaretti}
\affiliation[b]{Gran Sasso Science Institute, Viale Francesco Crispi 7, 67100 L'Aquila, Italy}
\author[c]{Francesco Vissani}
\affiliation[c]{Laboratori Nazionali del Gran Sasso, Via G. Acitelli 22, 67100 Assergi (L'Aquila), Italy}

\emailAdd{andrea.palladino@desy.de}
\emailAdd{carlo.mascaretti@gssi.it}
\emailAdd{francesco.vissani@lngs.infn.it}

\date{\today}

\abstract{The evidence of a new population of diffuse high-energy neutrinos, obtained by IceCube, 
has opened a new era in the field of neutrino physics. 
\andreamod{While most of the detected events are without any source counterpart, 
they are} compatible with the standard picture of
cosmic neutrinos undergoing 3-flavor neutrino oscillations. 
We analyze the implications of neutrino oscillations for the present and future experiments, 
focusing particularly on tau neutrinos. 
\andreafin{In fact tau neutrinos are very important}: even if they are not produced in astrophysical sites, they have to exist due to oscillations, and their observation should be regarded as a basic proof in support of this scenario. 
Moreover, IceCube's measurement of the flux of muon neutrinos implies that the flux of tau neutrinos is measured within $20\%$, just assuming standard neutrino oscillations. 
On this basis, after discussing the experimental signatures of tau neutrinos, we predict the rates for $\nu_\tau$ detection in the present and future neutrino telescopes. 
We show that \andreafin{the present IceCube detector is close to observe the first tau neutrinos, with a probability of about 90\%. Moreover} the next generation of \andreafin{IceCube} can identify about 2 neutrinos per year, reaching an evidence of 5$\sigma$ in about 10 years, despite the present uncertainty on the spectrum and on the production mechanism. 
The non observation of these neutrino events would have dramatic implications, such as the questioning of cosmic neutrino observations or the violation of neutrino oscillations over cosmological scales.}

\maketitle

\tableofcontents

\section{Introduction}
The discovery of a diffuse flux of high energy neutrinos by IceCube \cite{icescience} has opened a new era in the field of neutrino astronomy. 
The sources of these neutrinos are still unknown, several candidates are into the game. 
In the standard astrophysical scenario, where secondary particles are due mostly to pion and kaon decays, only electron and muon neutrinos are produced, while tau neutrinos emerge due to neutrino oscillations during their propagation along cosmic distances.\footnote{Note incidentally that the fact that neutrinos come in three flavors makes neutrino astronomy intrinsically multi-messenger.}
This natural possibility was stressed shortly after the discovery of the $\tau$ lepton; in fact, in \cite{bp} we read,
\begin{quote}
{\it If there exist more than two neutrino types with mixing of all neutrinos, cosmic neutrino oscillations may result in the appearance of new type neutrinos, the field of which may be present in the weak interaction hamiltonian together with heavy charged lepton fields.} 
\end{quote}
Different production mechanisms yield distinguishable flavor composition \cite{cons1} and $\nu_\tau$ fluxes on Earth (as we will see), even if the current low statistics does not permit us to test these predictions \cite{cons2,icecombined}; the residual uncertainties on the oscillation parameters, instead, are small.

Distinctive features of tau neutrinos are observable at very high energy, whereas below few hundreds of TeV they produce a shower-like event, similar to charged-current interactions of $\nu_e$ and to neutral current interactions of all neutrino flavors. 
The characteristic signature of tau neutrinos is the double vertex of interaction, i.e.~the vertex of tau lepton production (with hadronic interactions) and the subsequent vertex of tau decay. 
The observation of two vertices of interaction can be realized using two different techniques; the double bang \cite{pakvasatau} and the double pulse \cite{icetau}. 
We explain the difference between these two techniques in Sec.~\ref{sec:III}. 

At the time of the writing tau neutrinos are still not detected in IceCube. 
The expected number of signal events in six years of IceCube is $\sim 2.3$ for an $E^{-2.5}$ spectrum as in \cite{icrc}, which is still compatible with the non-observation of tau neutrinos within 90\% C.L.. On the other hand the non-observation of tau neutrinos can become an important issue in the near future, with the next generation of detectors. 
Increasing the volume of the detector from \SI{1}{\cubic\kilo\meter} to \SI{10}{\cubic\kilo\meter}, a tau neutrino \textit{must} be observed, \carlo{assuming standard oscillations}, in few years. 
As we will show in the paper, this consideration does not depend on the assumptions on the shape of the spectrum or on the assumption on the production mechanism, and, as a matter of fact, the present knowledge of astrophysical neutrinos is sufficient to make sufficiently robust predictions.

The non-observation of $\nu_\tau$'s in the future would imply dramatic consequences in neutrino oscillations and neutrino astronomy, because it would mean:\andreamod{
\begin{itemize}
\item violation of standard neutrino oscillations, that would imply new physics in the neutrino sector;
\item crisis of cosmic neutrino observations.
\end{itemize}}
For these reasons it is very important to quantify the current and future theoretical expectations on tau neutrino events. We evaluate the number of years required to observe a $\nu_\tau$ with a 5$\sigma$ confidence level. 
Just as the OPERA experiment, designed to prove the occurrence of oscillations, achieved a 5$\sigma$ evidence \cite{opera}, a reliable measurement of high energy $\nu_\tau$ would be a proof that the neutrinos seen by IceCube reach us from cosmic distances. 
We can refer to this important measurement as a ``cosmic OPERA'' experiment.

The structure of this work is the following: in Sec.~\ref{sec:I} we present a theoretical overview of neutrino oscillations, that can be read as the present \lq\lq state of the art\rq\rq; in Sec.~\ref{sec:II} we estimate the fraction of $\nu_\tau$ at Earth, considering different production mechanisms at the source; in Sec.~\ref{sec:III} we present the methods that are used to obtain the effective areas for different detectors, starting from first principles related to the physical processes. 
We use these results to derive predictions for tau neutrino events, and we present the results in Sec.~\ref{sec:IV}. 
We conclude the work with a discussion in Sec.~\ref{sec:V}. 

\section{Theoretical overview}
\label{sec:I}
Non-zero neutrino masses have been considered seriously  
\carlo{since the neutrino was proposed} (Pauli, Fermi, Perrin) and allow us to explain a wide set of phenomena.
While they imply modifications of the accepted standard model of elementary particles, these modifications fit well the conventional three family picture.  
Neutrino masses emerge quite naturally \cite{pm,wein}; if we describe them with effective operators \cite{wein}, neither the gauge group nor the particle content should be modified.  
This position is completely consistent with the currently available data and maintains predictive power. 
E.g., a transition magnetic moment is predicted to exist, even if it is expected to be small 
$\mu_\nu\sim 10^{-24} \mu_{\mbox{\tiny B}}$  \cite{shrock} (the adimensional constant 
$\sim 3 G_F m_e m_\nu /(16\sqrt{2}\pi^2) (m_\tau/m_W)^2$
is evaluated with $m_\nu\sim \sqrt{\Delta m^2}= 50\mbox{ meV}$, where $\Delta m^2=2.5\times 10^{-3}$ eV$^2$).
On the contrary, we do not have convincing theoretical indications in favor of other light neutrinos besides the usual three.\footnote{The same is largely true for the known phenomenology \cite{cirelli,book}. There are several anomalies that could be explained individually invoking new oscillations, however the overall picture lacks of consistency. Conversely, $Z$-width measurements rule out other interacting neutrinos species; big-bang nucleosynthesis is consistent with three neutrino species; the study of the anisotropies of the microwave distribution at cosmic scales indicates the same and together with cosmological measurements at smaller scales yields a tight bound on neutrino masses.}

The observed phenomenology of oscillations (see \cite{exp} for an incomplete list) can be explained with three massive neutrino states with different masses, which imply two different frequencies of oscillation in vacuum: these have been measured, testing the oscillation phases $\varphi_{ij}=(E_{\nu_i}-E_{\nu_j}) t/\hbar$ with $E_{\nu_i}=\sqrt{\mathbf{p}^2+m_{\nu_i}^2}$.
Solar neutrinos offer evidence for another effect for neutrinos that propagate in ordinary matter \cite{msw}, which implies an additional frequency of oscillation associated to the energy $E_{\nu_e}' -E_{\nu_\mu}' = E_{\nu_e}' -E_{\nu_\tau}'=\sqrt{2}G_F\, n_e$, where $n_e$ is the electronic number density. 

The importance of neutrino oscillations for cosmic neutrinos has been generally appreciated after \cite{pakvasatau}, even if their relevance and implications have been discussed even much in advance, see, e.g.,~\cite{venya,bp}. 
 The conventional formalism is the simplest one, namely averaged vacuum oscillations \cite{bp}. 
A particularly convenient parametrization of the effects of vacuum oscillations was discussed in \cite{natpar}. 
 Before passing to the quantitative considerations, we assess the reliability of the above theoretical setup.

\paragraph{Possible deviations from the minimal assumption}
\andreamod{The distances traveled by cosmic neutrinos are much larger than those explored 
directly and, moreover, the corresponding energies are very small, 
$\Delta m^2_\text{atm}\ L/(2 E_\nu^\star)= 6\times  10^{-17}\mbox{ eV }$
for  $\Delta m^2_\text{atm}= 2.5\times 10^{-3} \mbox{\,\small eV}^2$ and 
${E_\nu^\star}=50$ TeV.  These considerations suggest that 
new physics could play a role. Therefore, we examined a number of 
concrete possibilities:
\begin{enumerate}
\item The matter effect \cite{msw} could be relevant if neutrinos are produced in a dense medium. 
This was studied in \cite{razz}, finding that some effects exist for energies $\ll E_\nu^\star$, smaller than those of our interest.
\item If neutrinos are produced near a neutron star, the neutrino magnetic energy, $\delta H=- \boldsymbol{\mu}_\nu \cdot \mathbf{B}$ 
is of the same size as the vacuum term $\Delta m^2_\text{atm}\ L/(2 E_\nu^\star)$ 
for extreme parameters $B=10^{11}$~T and $\mu_\nu = 4\times 10^{-24} \mu_{\mbox{\tiny B}}$.
However, neutrino leave the star at velocity $c$, and the accumulated phase $\delta \varphi=T\,\delta H\sim 10^{-5}$ is very small. 
\item Suppose that different neutrinos propagate differently in gravitational fields \cite{gasp} having peculiar dispersion relations such 
as $E_{\nu_i}^2=\mathbf{p}^2+m_{\nu_i}^2+ \eta_{\nu_i} \mathbf{p}^4/M_{\mbox{\tiny Pl}}^2$ (where
is the Planck mass and $\eta_{\nu_i}$ are $i$-dependent coefficients)
see for a recent discussion \cite{gasp2}.
This would imply an effect for energies above $E_\nu^\star$ if $\eta_i-\eta_j\ge 1/20$; 
however, the theoretical motivations and the completeness of the modeling seems to be weak.
\item Certain models for sterile neutrinos, which would lead to new effect on cosmic scales, have slightly stronger  
theoretical bases.
These include the model with exact mirror symmetry  \cite{mirror} or pseudoDirac neutrinos \cite{pseudodirac};
in both cases, tau neutrinos have to be produced due to oscillations, just as in the three flavor case.\end{enumerate}}

\paragraph{Assessment of the vacuum oscillation hypothesis}

The above discussion does not strive to completeness but simply to argue the following position: even if in principle the propagation of cosmic neutrinos could be possibly affected by new effects, as (some of) those examined above or other ones - for instance, those described in~\cite{cgt,Rasmussen:2017ert} - we fail to see any convincing reason at present to assume a deviation from the conventional picture. In particular there is {\em no such indication from the available data from IceCube, interpreted in terms of cosmic neutrinos}, see e.g., 
\cite{cons1,cons2,cons3}.
Conversely, in the most plausible astrophysical situations, the sites of neutrino production are typically almost empty and the simple and assessed picture of neutrino propagation, that includes only three flavor vacuum oscillations, does apply.

\begin{figure}[t]
\centering 
\includegraphics[width=0.6\textwidth]{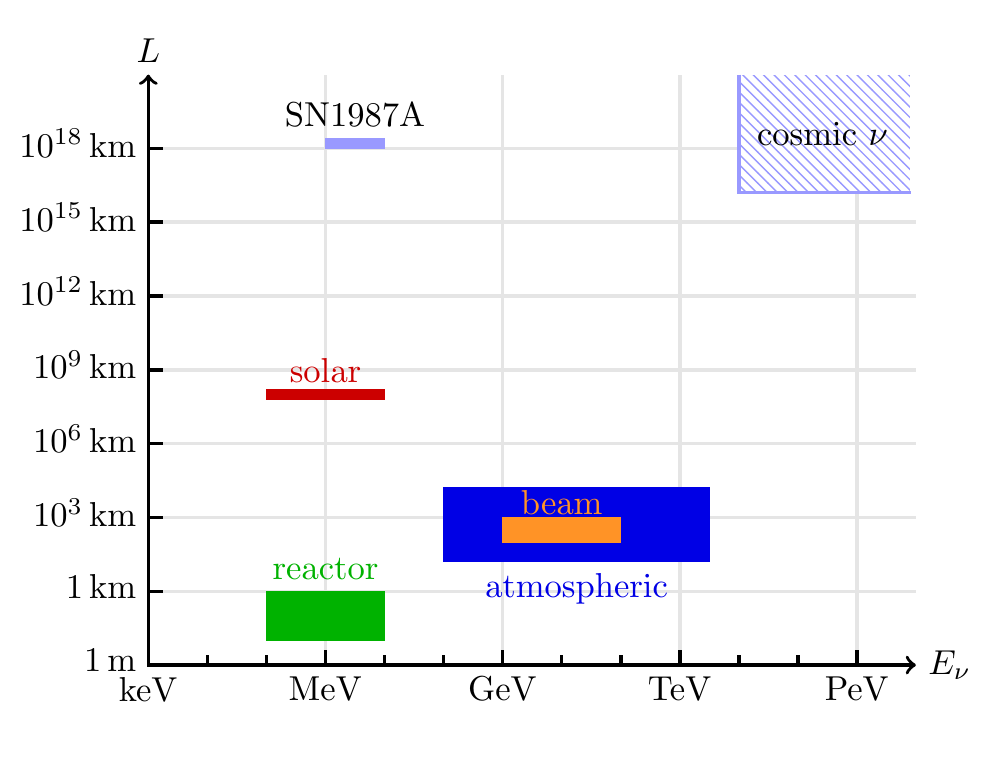}
\caption{The regions of the $L$ and $E_\nu$ space relevant for experiments on neutrino oscillations (red, blue, orange and green rectangles), for SN1987 (light blue rectangle) and for the neutrinos that can be detected in IceCube (light blue shaded rectangle).}
\label{fig:nu_oscs}
\end{figure}

As a final observation apropos, we would like to \carlo{remind} 
that the ratio $L/E_\nu$, relevant for the vacuum oscillations phase of high energy neutrinos, is similar to other $L/E_\nu$ ratios that have been already probed with low energy neutrino astronomy. 

In view of these reasons, we will proceed
with a detailed quantitative exploration of the minimal hypothesis that concerns the propagation of cosmic neutrinos, namely, we will assume the occurrence of three-flavor vacuum  oscillations and no other phenomena.


\paragraph{Oscillation and survival probabilities}
The distances and the energies for several interesting cases, relevant for  vacuum neutrino oscillation studies, are resumed in Fig.~\ref{fig:nu_oscs}. 

In the case in which we are interested (ordinary three flavor neutrinos) the phases of propagation are very large.
Therefore, the values of the three-flavor oscillation/survival probabilities  can be written as \cite{bp}:
\[P_{\ell\ell'} = \sum_{i=1}^3 |U_{\ell i}^2 | |U_{\ell' i} ^2| \qquad \ell,\ell' = e,\mu,\tau \]

\begin{figure*}[t]
\centering 
\includegraphics[width=1\textwidth]{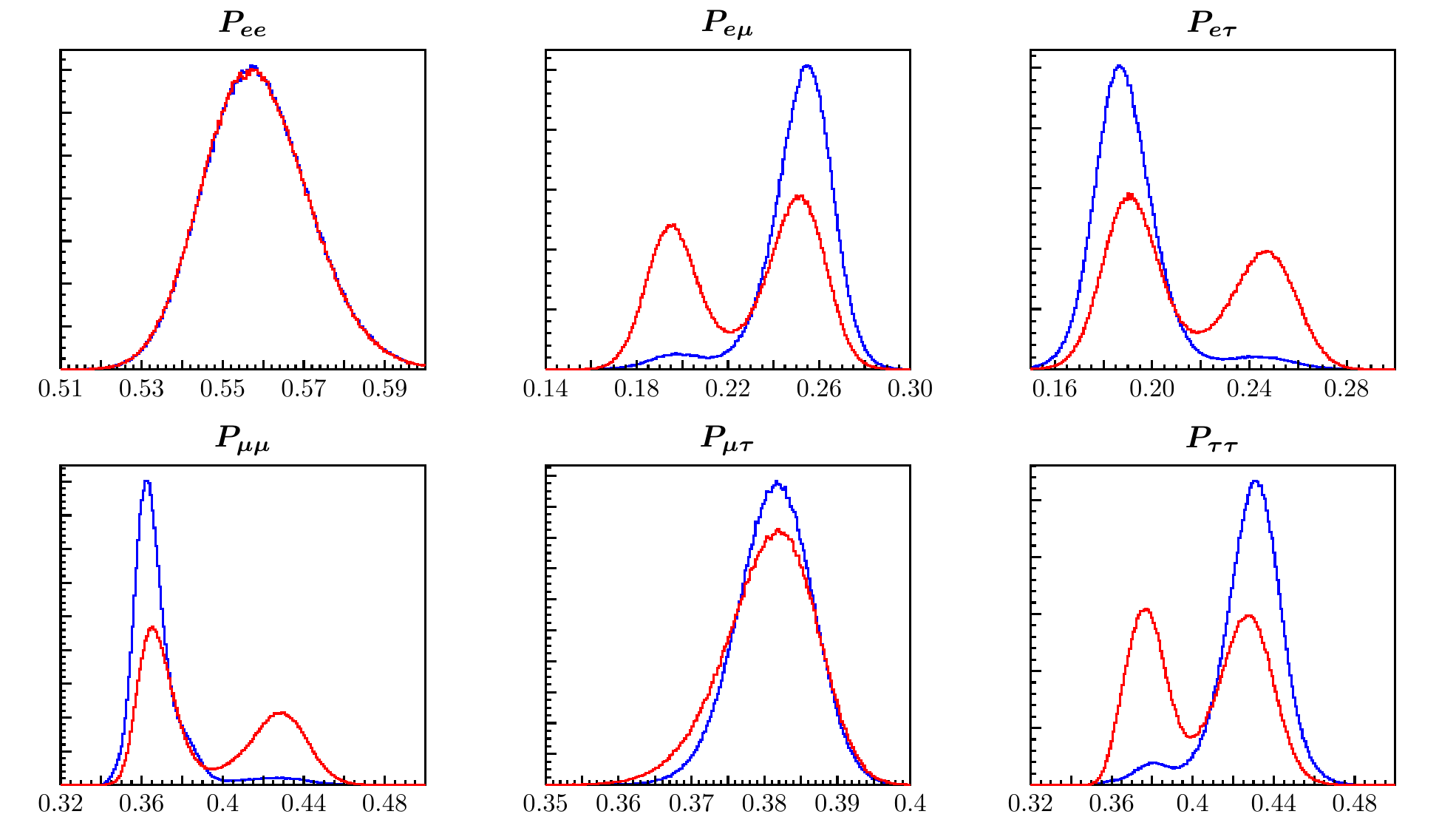}
\caption{The PDFs of the oscillation/survival probabilities $P_{\ell\ell'}$ in the case of normal ordering (blue) and inverse ordering (red), in the average regime.}
\label{fig:pll'}
\end{figure*}

We computed the distributions of $P_{\ell\ell'}$ starting from the best fit to the neutrino oscillation experiments by \cite{bari}. 
Sampling the oscillation parameters distributions of Fig.~1 of \cite{bari}, we obtained the distributions for the $P_{\ell\ell'}$, which we show in Fig.~\ref{fig:pll'}.

\section{Tau neutrinos from neutrino oscillations} \label{sec:II}
\begin{figure*}[t]
\centering
\includegraphics[width=0.31\textwidth]{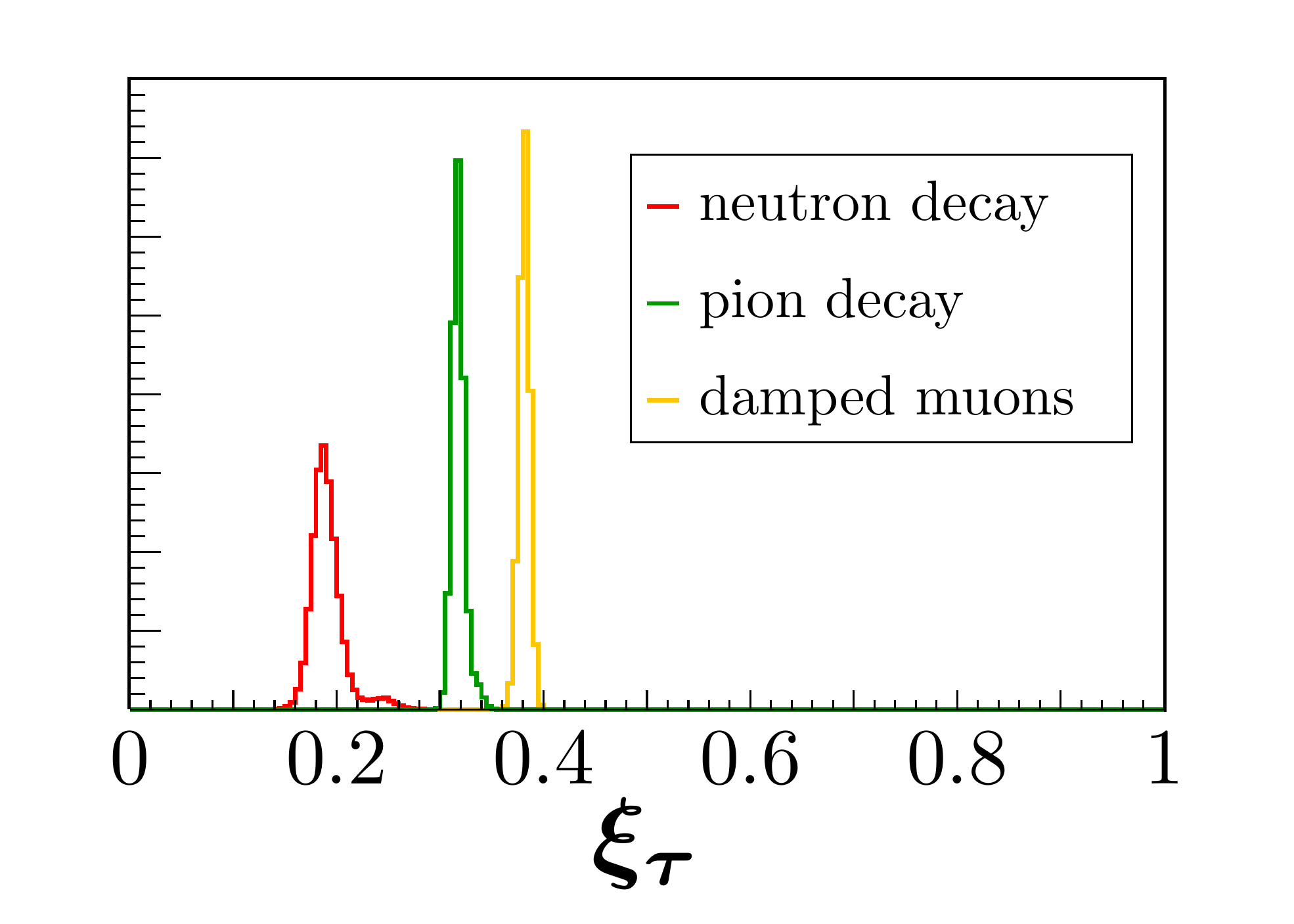}
\includegraphics[width=0.33\textwidth]{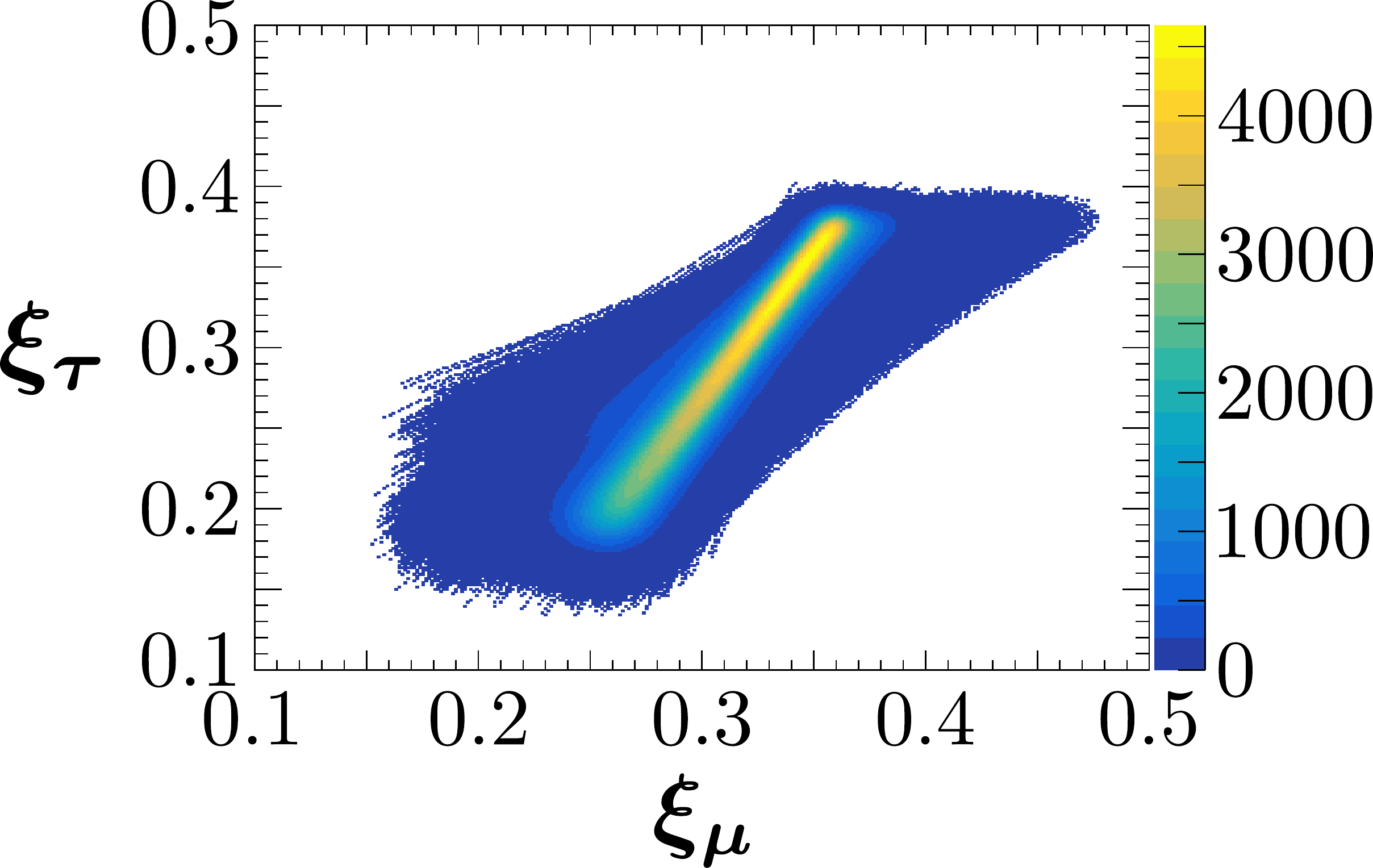}
\includegraphics[width=0.32\textwidth]{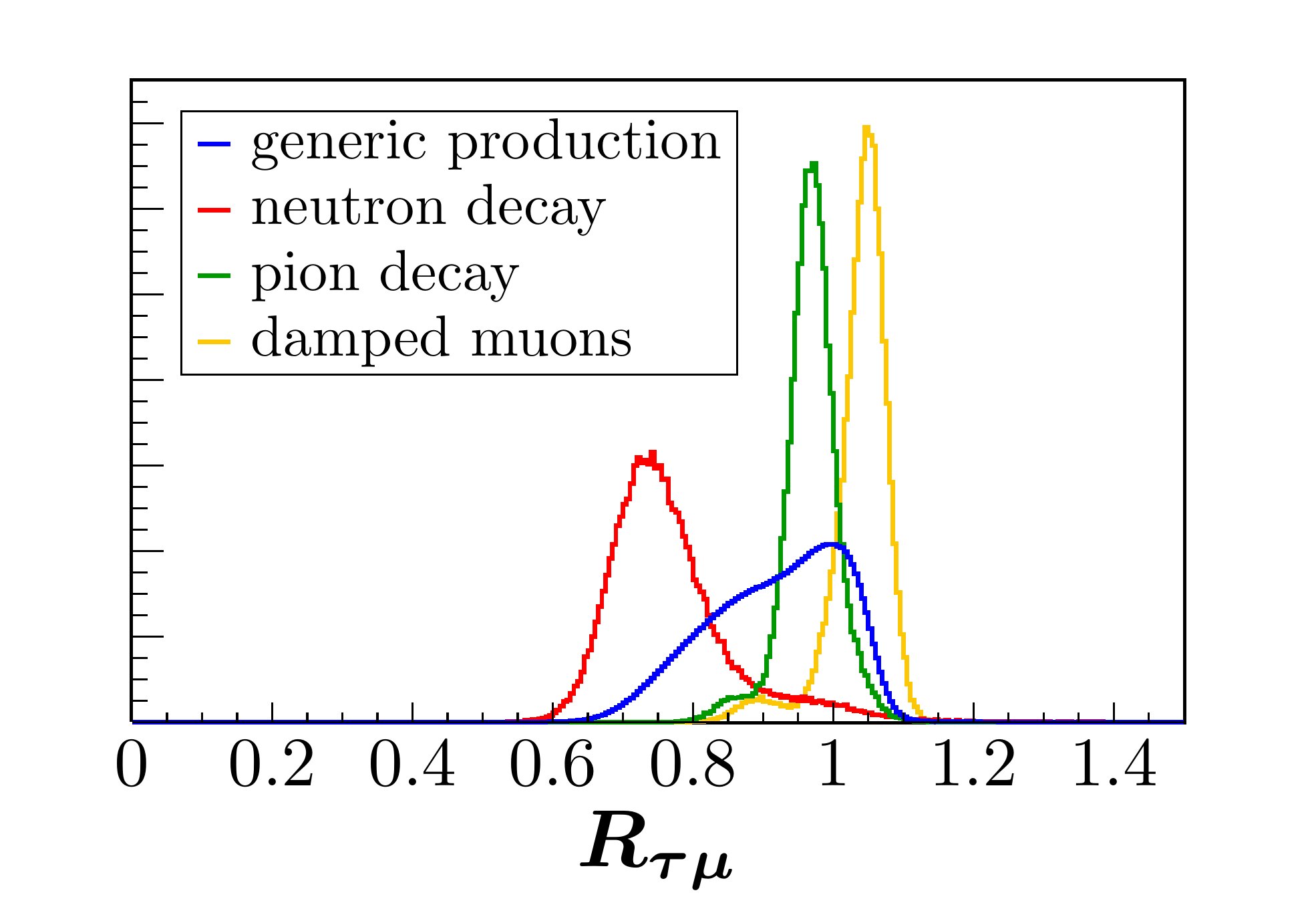}
\caption{Left panel: the PDFs of the fraction of tau neutrinos at Earth expected from standard production mechanisms of high energy neutrinos, after neutrino oscillations. 
Central panel: the astrophysical tau neutrino flux fraction at Earth $\xi_\tau$ vs.~the astrophysical muon neutrino flux fraction at Earth $\xi_\mu$. 
\carlo{The color bar indicates how many times (out of $10^7$) a certain bin has been filled.}
Right panel: the PDFs of the ratio between the flux of tau neutrinos and flux of muon neutrinos (at Earth) for different production mechanisms and for a generic situation (blue line). 
These plots have been obtained sampling the survival/oscillation probabilities according to their distribution about $10^5$ times for each $\xi_\mu^0$, which has been picked uniformly between 0 and 1.
\carlo{For these three plots, and from now on, we assume normal ordering.}}
\label{fig:taufrac}
\end{figure*}

Tau neutrinos are assumed not to be produced in standard astrophysical mechanisms of high energy neutrino production. 
On the other hand, tau neutrinos are always expected at Earth due to \carlo{standard} neutrino oscillations, as shown in Fig.~\ref{fig:taufrac}.

As shown in Fig.~\ref{fig:taufrac} (discussed below) and consistently with \cite{cons3}, the observation of an astrophysical muon neutrino flux \cite{icescience,icemuon} by itself would imply a very similar tau neutrino flux at Earth, assuming the validity of standard three-flavor neutrino oscillations up to these energies.

By considering the flavor fractions before and after oscillations, 
$$
\xi_\ell =\frac{\Phi_{\nu_\ell}}{\Phi_{\nu_e} + \Phi_{\nu_\mu}+\Phi_{\nu_\tau}}\ , \ \xi_\ell^0 =\frac{\Phi_{\nu_\ell}^0}{\Phi_{\nu_e} + \Phi_{\nu_\mu}+\Phi_{\nu_\tau}}
$$
where $0\le \xi_\ell \le 1$, $0\le \xi_\ell^0 \le 1$ and $\ell=e,\mu,\tau$, 
we obtain the astrophysical neutrino flux fraction at Earth after oscillations, 
\begin{equation}
\xi_\ell = P_{\ell e} (1-\xi_\mu^0 -\xi_\tau^0 ) + P_{\ell \mu}\xi_\mu^0 + P_{\ell \tau} \xi_\tau^0 
\label{eq:flavor}
\end{equation}
where  the generic mechanism of production at the source is described by two parameters,
\[ (\nu_e:\nu_\mu:\nu_\tau) = (1-\xi_\mu^0-\xi_\tau^0 :\xi_\mu^0:\xi_\tau^0)  
\] 
subject to the condition $0\le 1-\xi_\mu^0-\xi_\tau^0\le 1$. 
Starting from this general flavor composition at the source, it is possible to evaluate the ratio between the flux of $\nu_\tau$ and the flux of $\nu_\mu$ at Earth, that we call $R_{\tau\mu}$. 
This quantity is particularly interesting since the flux of $\nu_\mu$ is measured by means of the throughgoing muon flux. 
The function $R_{\tau\mu}$ is given by the following expression:
$$
R_{\tau\mu}= \frac{P_{e\tau} + \xi_\mu^0 (P_{\mu\tau}-P_{e\tau})+ \xi_\tau^0(P_{\tau\tau}-P_{e\tau})}{P_{e\mu} + \xi_\mu^0 (P_{\mu\mu}-P_{e\mu})+ \xi_\tau^0(P_{\mu\tau}-P_{e\mu})}
$$

In standard astrophysical environments tau neutrinos are not produced at the source, that motivates us to assume $\xi_\tau^0=0$. 
If a small fraction of $\nu_\tau$ is present at the source, the flux of $\nu_\tau$ at Earth would slightly increase.

Under this assumption the previous equation becomes simpler:
\begin{equation}
R_{\tau\mu}= \frac{P_{e\tau} + \xi_\mu^0 (P_{\mu\tau}-P_{e\tau})}{P_{e\mu} + \xi_\mu^0 (P_{\mu\mu}-P_{e\mu})}
\label{eq:rtaumu}
\end{equation}
In the rest of this work, we will consider the expression for $R_{\tau\mu}$ given in eq.~\eqref{eq:rtaumu}; now, we proceed to quantify its value.

In the left panel of Fig.~\ref{fig:taufrac}, \carlo{we show} the expected fraction
 of $\nu_\tau$ at Earth, using Eq.~\eqref{eq:flavor} and 
 taking into account the uncertainties on neutrino oscillations. 
The production mechanisms at the source that are considered are the following:
\begin{itemize}
\item $(\xi_e^0:\xi_\mu^0:\xi_\tau^0)=(1:0:0)$, neutron decay;
\item $(\xi_e^0:\xi_\mu^0:\xi_\tau^0)=(1/3:2/3:0)$, pion decay;
\item $(\xi_e^0:\xi_\mu^0:\xi_\tau^0)=(0:1:0)$, damped muons.
\end{itemize}
The second is the most plausible one (possibly, with minor variants), 
and the other two are introduced mostly for the purpose of comparison.

Sampling $\xi_\mu^0$ uniformly in $[0,1]$ and sampling the oscillation parameters $P_{\ell\ell'}$ according to their distribution (see Fig.~\ref{fig:pll'}), we obtain the second panel of Fig.~\ref{fig:taufrac}, while the quantity relevant for the rightmost panel of Fig.~\ref{fig:taufrac} is the ratio of the tau neutrino flux to the muon neutrino flux at Earth, defined in Eq.~\eqref{eq:rtaumu}. Using the normal hierarchy in the case of $\xi_\mu^0 \in [0,1]$, we obtain 
$$
R_{\tau\mu}^{\text{NH}}=1.00^{+0.05}_{-0.15}
$$
and its distribution is represented by the blue line in the rightmost panel of Fig.~\ref{fig:taufrac}.

We notice that the flux of $\nu_\tau$ is strictly related to the flux of $\nu_\mu$: 
$$\phi_\tau \geq 0.78 \, \phi_\mu \ \ \ \mbox{within 90\% C.L.}$$
This means that once $\phi_\mu$ is measured, $\phi_\tau$ is very strongly constrained, if standard neutrino oscillations hold. 
This consideration is relevant, since $\phi_\mu$ is measured in IceCube by means of the throughgoing muon signal, i.e.\ the flux of muons generated by $\nu_\mu$ coming from the hemisphere opposite to the one in which the neutrino telescope is located. 
In the case of IceCube the throughgoing muons come from the Northern hemisphere \cite{icemuon}, whereas for the \carlo{upcoming} KM3NeT they will come mostly from the Southern hemisphere.

The previous results are obtained using the normal hierarchy, that is favored at approximately $2\sigma$ by \cite{bari}. 
Using the inverted hierarchy in the case of $\xi_\mu^0 \in [0,1]$, we obtain 
$$R_{\tau\mu}^{\text{IH}}= 1.02^{+0.04}_{-0.15} $$
The quantity $R_{\tau\mu}$ is relevant, since it will be used in Sec.~\ref{sec:IV} to predict the expected number of double cascades in different detectors.

\section{Double pulse and double bang: the effective areas} \label{sec:III}
As stated in the introduction, a tau neutrino usually produces a shower-like event and it cannot be distinguished from other neutrinos: in fact, a $\nu_e$ that interacts via charged current interaction, or whatever neutrino that interacts via neutral current interaction, would produce the same shower-like event. 

A tau neutrino can be unequivocally identified when two vertices of interaction become visible, and this happens in the so-called double bangs and double pulses. \andreamod{Before discussing this kind of events we need to introduce some concepts related to the detector. The neutrino telescopes are characterized by several digital optical modules (DOMs) enclosed in strings. There are 86 DOMs for each string in IceCube, but this number can be different for other experiments.
The separation between the strings is a peculiar feature of the detector and it is 120 meters in IceCube \cite{icescience}, 240 meters in IceCube-gen2 \cite{ice2} and 90 meters in KM3NeT~\cite{km3netloi}. 

Now we can explain the difference between double pulse and double bang:}
\begin{itemize}
\item a double bang consists in the observation of two vertices of interaction in two different strings, which are separated by $\sim$ 100 meters in the neutrino telescopes of our concern. 
This kind of events can be produced only by tau neutrinos having multi-PeV energy. 
This was the historical method proposed in the past to observe directly a $\nu_\tau$ \cite{pakvasatau};
\item a double pulse event corresponds to the detection of two subsequent signals in the same DOM. 
In this case the tau lepton has to travel few tens of meters in order to produce two distinguishable vertices of interaction. 
For this reason, only tau neutrinos of few hundreds of TeV are capable to produce such an event \cite{icetau}. 
\end{itemize}
\andreamod{In both cases the vertices of interaction must be contained into the detector, therefore double bangs and double pulses are subclasses of the so-called \lq\lq High Energy Starting Events\rq\rq \ (HESE), where \lq\lq Starting\rq\rq \ means that the vertex of interaction is contained into the detector. On the contrary, the case of $\nu_\mu$ that interact via charged current outside the detector is called \lq\lq throughgoing muon\rq\rq; we will also refer to this class of events in the following of this work, since this class of events can provide important information on the flux at very high energy (above 200 TeV).}

\andreamod{From here on} we will also use the generic term \lq\lq double cascade\rq\rq \ when we consider the two processes together. A general parametrization for the double cascade effective area, which is appropriate for our purposes and that it is based on the analytical expression proposed in \cite{palltau}, is given by the following formula:
%
%
%
\begin{equation}
A_{\text{eff}}(E_\nu,E_{\text{min}})=\frac{\rho V}{m_n} \,  \frac{1+ S(E_\nu)}{2}\times \text{BR}  \times
 \sigma_{\text {cc}}(E_\nu) \, \exp\left(-\frac{E_{\text {min}}}{E_\nu}\right)
\label{eq:aeff}
\end{equation}
where $\rho$ is the density of the material, $V$ the volume of the detector, $m_n$ the nucleon mass. 
The function $S(E_\nu)$ is the probability for neutrinos to cross the Earth, namely 0.91, 0.66, 0.37 and 0.18 at 10, 100, 1000 and 10000 TeV \cite{palltau}.
The parameter $\text{BR} \simeq 80\%$ denotes the branching ratio of the hadronic decay modes of the tau, which allow the second cascade to be visible (basically it excludes the channel in which a tau lepton decays into a muon \cite{pdg}). 
The function $\sigma_{\text {cc}}(E_\nu)$ is the charged current cross section \cite{Gandhi:1998ri} (averaged for neutrinos and antineutrinos) and $E_{\text {min}}$ is the minimum energy required to observe two vertices of interactions. 
As stated before, this energy is order of sub-PeV for double pulses and multi-PeV for double bangs; for this reason, double pulse events are intrinsically more likely to detected than double bang events.

\begin{figure*}[t]
\centering
\includegraphics[width=0.32\textwidth]{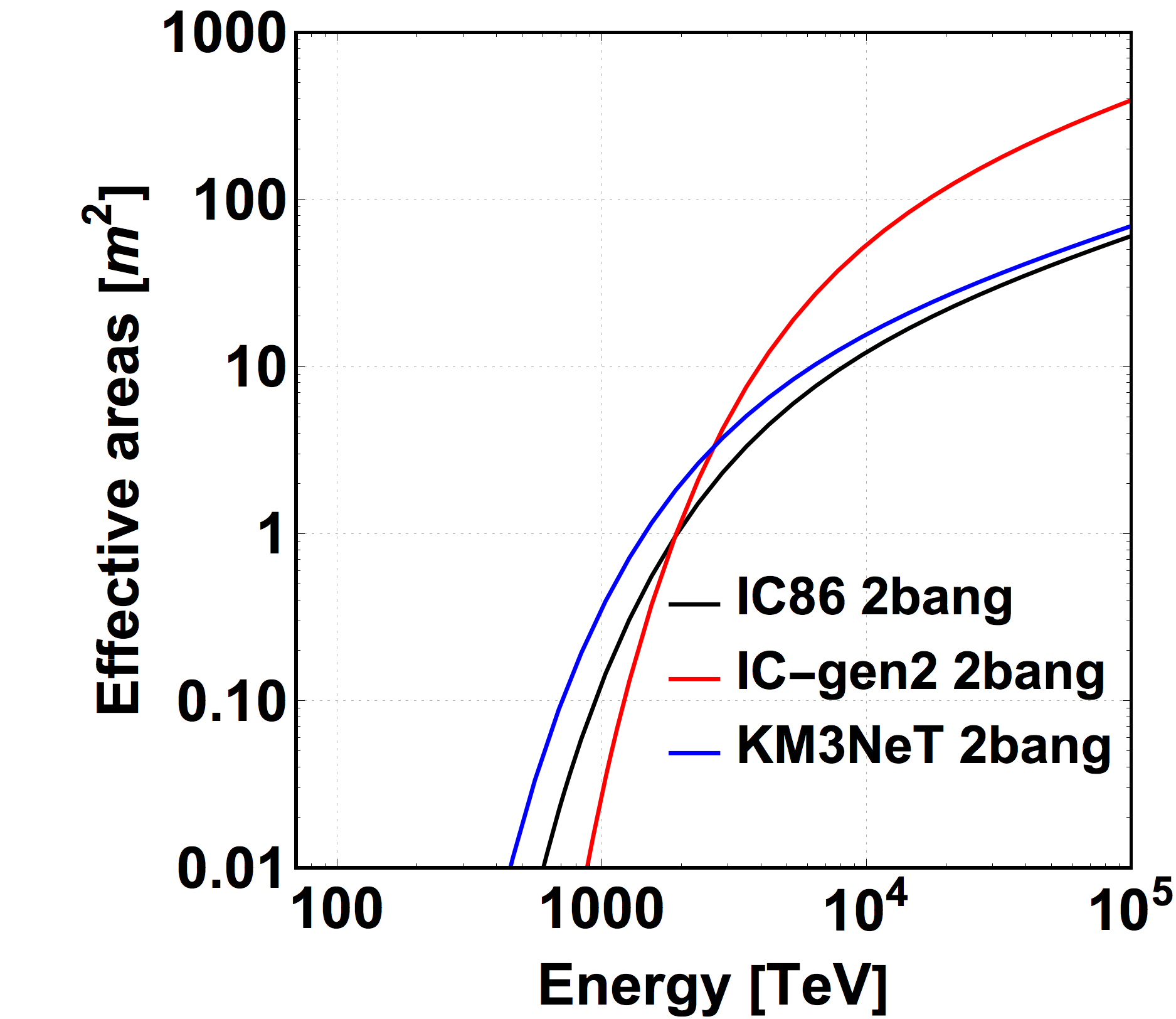}
\includegraphics[width=0.32\textwidth]{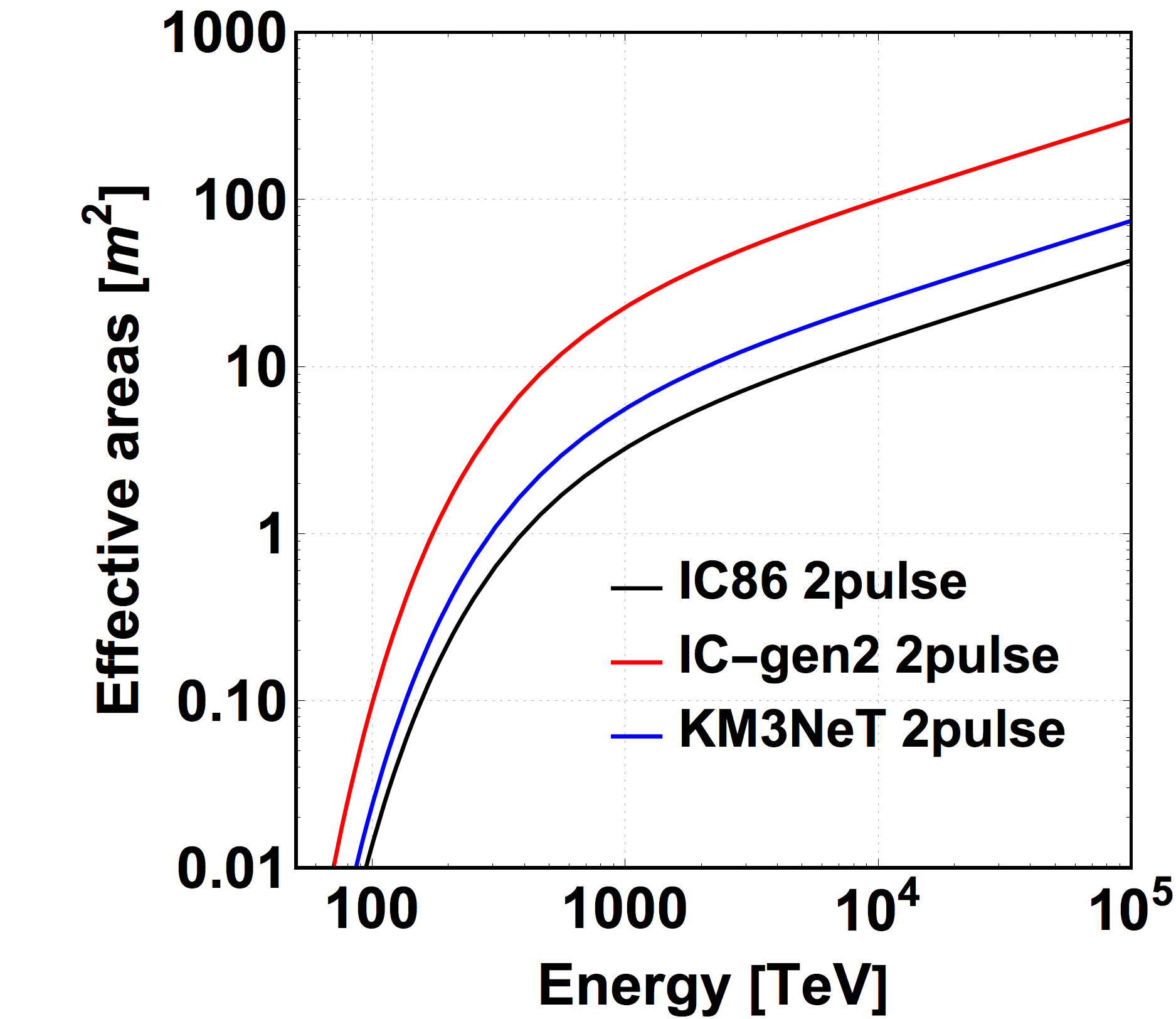}
\includegraphics[width=0.32\textwidth]{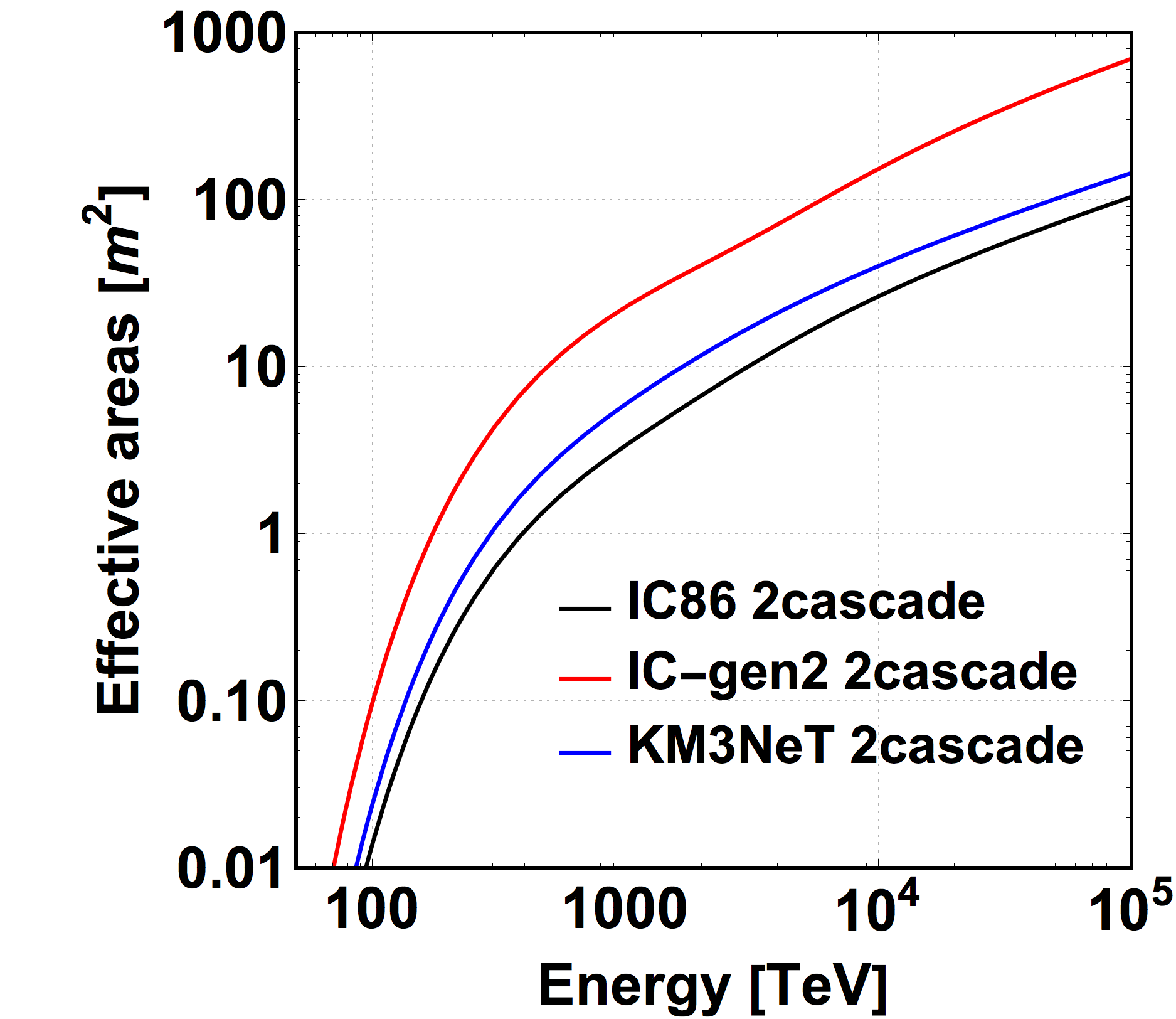}
\caption{Simulated effective areas of double bang, double pulse and double cascade (double pulse + double bang) events for present and future neutrino telescopes.}
\label{fig:aeff}
\end{figure*}

\subsection{Double bangs}
In this subsection we examine the first method that was proposed to observe tau neutrinos, i.e.~the so called double bang \cite{pakvasatau}, showing how it is possible to obtain an estimation of the effective area starting from theoretical considerations of the relevant physics. \andreamod{(Before proceeding, note that at the energies of interest for the current IceCube dataset, $\nu_\tau$ and $\tau$ energy losses are negligible, see footnote \ref{fooo} for a discussion.)}

\andreamod{It is possible to estimate the minimum energy required for a $\nu_\tau$ in order to produce two distinguishable signals in two different strings. In order to do that we take into account the lifetime of the $\nu_\tau$.} 
It is easy to verify that a tau neutrino of 1 PeV travels $\sim$ 50 meters before decaying. 
Moreover, the energy of the tau lepton is about 3/4 the energy of the primary $\nu_\tau$.
Combining these conditions, we can compute the (average) minimum energy as a function of the required minimum length as follows:
\begin{equation}
E_{\text{min}}= \frac{4}{3} \left(\frac{d}{50 \mbox{ m}}\right)\,\mbox{PeV} 
\label{eqemin}
\end{equation}
where $d$ is the distance between the strings.
Therefore, the minimum energy is equal to 2.4 PeV, 3.2 PeV, and 6.4 PeV for KM3NeT, IceCube and IceCube-gen2, respectively.  

The effective area of double bang events is not published, but it is possible to estimate it starting from the effective area of tau neutrinos given in \cite{icescience}.
Comparing our parametrization given in Eq.~\eqref{eq:aeff} (including also the neutral current cross section, that is included in the total effective area of $\nu_\tau$ reported in \cite{icescience}) we found that, far above the 30 TeV energy threshold, the effective area of $\nu_\tau$ is well reproduced (with coefficient of determination $R^2=0.99$) using an effective volume $V=0.97 \ \mbox{km}^3$ and a minimum energy of 100 TeV.\footnote{The meaning of this minimum energy is not relevant for the treatment of double bangs and double pulses. Anyway, in order to check the plausibility of this calculation, let us recall that the energy threshold of IceCube High Energy Starting Events (whose the effective area of $\nu_\tau$ is connected) is 30 TeV of deposited energy. The deposited energy of a tau neutrinos is about 70\%-80\% in charged current interaction and 25\% in neutral current interaction (see \cite{pallwinter} in which the connection between the deposited energy and the reconstructed energy is discussed). Therefore, \andreamod{a minimum incident energy of 100 TeV would correspond to a deposited energy between 20-80 TeV, that is reasonably in agreement with the energy threshold of 30 TeV, that characterizes the high energy starting events (HESE).}}
This means that the effective volume is similar to the physical volume of the detector for tau neutrinos.

In order to estimate the double bang effective area we need to consider the minimum energies described before, in order to cover the distance between two strings.
Moreover, we have to consider that the path necessary to reach the second string is a function of the incident angle of neutrinos. \andreamod{In fact, the distance between two different strings is equal to $d/\cos\theta$, where $\theta$ denotes the angle between the neutrino's direction and the plane perpendicular to the strings. Therefore the minimum energy required for a $\nu_\tau$ will be greater than $E_{\text{min}}$ for $\theta \neq 0$, namely $E_{\text{min}}/\cos\theta$ following \eqref{eqemin}}.
Taking into account this aspect, we obtain the general parametrization of the double bang effective area as follows:
\begin{equation}
A_{\text{eff}}^{\text{2bang}}(E_\nu,E_{\text{min}})=\frac{1}{\pi}\int_{-\pi/2}^{\pi/2} A_{\text{eff}}\left(E_\nu,\frac{E_{\text{min}}}{\cos\theta} \right) d\theta
\end{equation}
The previous formula is valid for an isotropic flux of neutrinos, as expected from cosmic neutrinos.
In the left panel of Fig.~\ref{fig:aeff} we report the simulated effective areas for double bangs for the three different neutrino telescopes that we have considered. 
We denote the three experiments with the name IC86 (IceCube), IC-gen2 (IceCube generation 2) and KM3NeT. 
The parameters used to obtain them are reported in Tab.~\ref{default}. 
For IC-gen2 we consider that the new detector will have a sensitivity about 7 times larger than the present IceCube \cite{ice2}, therefore the effective volume for the double bang detection would be $V=6.8 \ \mbox{km}^3$. 
For KM3NeT, instead, we limit our analysis to the ideal scenario of $V=1 \ \mbox{km}^3$, since there is still no information available for this process in this future neutrino telescope. 
In view of the fact that the effective volume of double bang for IceCube is close to \SI{1}{\cubic\kilo\meter}, we believe that our approximation is adequate.

\begin{table}[t]
\caption{Table of the parameters related to the double pulse and double bang effective areas for IceCube, IceCube-gen2 and KM3NeT. The parameters for IceCube are based on the present detector whereas the parameters for IceCube-gen2 and KM3NeT are estimated.}
\begin{center}
\begin{tabular}{llcccc}
\midrule
\multirow{2}{*}{Detector} &\multirow{2}{*}{Event type} &$\rho$ &$V$ &$E_{\text{min}}$ &$d$\\
&&(\si{\gram\per\cubic\centi\meter}) &(\si{\cubic\kilo\meter}) &(\si{\peta\electronvolt}) &(\si{\meter})  \\
\midrule
IC86 & Double pulse & 0.917 & 0.63 & 0.58 & \phantom022 \\
IC86 & Double bang & 0.917 & 0.97  &  3.2\phantom0 &  120 \\
IC-gen2 & Double pulse & 0.917 & 4.4\phantom0 & 0.58 & \phantom022 \\
IC-gen2 & Double bang & 0.917 & 6.8\phantom0  &  6.4\phantom0 &  240 \\
KM3NeT& Double pulse & 1& 1\phantom{.00} &  0.58 & \phantom022\\
KM3NeT & Double bang & 1 & 1\phantom{.00} &  2.4\phantom0 & \phantom090 \\
\midrule
\end{tabular}
\end{center}
\label{default}
\end{table}%

\subsection{Double pulses}
In this subsection we focus on double pulses, i.e.\ the processes in which the two vertices of interaction are identified in the same optical module. 
Comparing \eqref{eq:aeff} with the IceCube double pulse effective area given in \cite{icetau}, as already done in \cite{palltau}, we found that it is well reproduced (within an average uncertainty of 5\% between 100 TeV and 10 PeV) by the set of parameters $V=0.28 \ \mbox{km}^3 \ E_{\text{min}}=0.5 \ \mbox{PeV}$. This means that the effective volume for this kind of event is a factor 3.5 smaller than the physical volume of the detector. 
Let us recall that in \cite{icetau} the expected number of double pulses in 4 years was $\sim 0.5$ events, assuming an $E^{-2}$ spectrum.

Recently IceCube has presented an updated (preliminary) analysis of the double cascade events expected in 6 years in \cite{icrc}. The expected number of identifiable astrophysical tau neutrinos is claimed to be equal to:
$$
N_{\text{2p}}^{\text{IC86}}=2.318^{+0.038}_{-0.029}
$$
after 6 years of exposure, considering an $E^{-2.3}$ spectrum, with normalization at 100 TeV of $1.5 \times 10^{-18}\,\text{GeV}^{-1} \text{cm}^{-2} \text{sec}^{-1} \text{sr}^{-1}$. 
We denote this flux by $d\phi/dE_\nu$. 
In this case the effective area has not been released, so that the best we can do is to use the same minimum energy of the previous analysis, changing the effective volume in order to reproduce the expected number of events.

The expected number of events can be computed using the following formula:
$$
N_{\text{2p}}(E_{\text{min}})=4\pi \text{T} \int_0^\infty \frac{d\phi}{dE_\nu} \, A_{\text{eff}}(E_\nu,E_{\text{min}}) \, dE_\nu
$$

Assuming that the expectations reported in \cite{icrc} are related to double cascade events (double pulse + double bang), 
in order to be in agreement with them the effective volume of Eq.~\eqref{eq:aeff} must be equal to $V=0.63 \ \mbox{km}^3$, fixing $E_{\text{min}}=0.5$ PeV, as in the previous analysis (this value, in fact, depends on the features of the process, not on the optimization of the analysis).
Under this assumption\footnote{Let us remark that it is a conservative hypothesis, because the assumption that 2.3 events are created only by double pulses would increase the expected number of total double cascade events, when also the contribution of the double bang events is taken into account. This would make our conclusions even stronger. 
}, a fraction of the expected number of events is provided by double bangs. 

In our understanding, this conclusion means that IceCube has performed an optimization of the analysis dedicated to the research of tau neutrinos, gaining more than a factor 2 in the expected number of events when compared to the previous analysis \cite{icetau}. Summarizing, the effective area for double pulse events is exactly given by the same formula of Eq.~\eqref{eq:aeff}.
In Fig.~\ref{fig:aeff} we report the simulated effective areas for the future neutrino telescopes, obtained with the parameters reported in Tab.~\ref{default}. 

The search for tau neutrinos with energies larger than the maximum currently observed in IceCube, $\sim 5$ PeV, is also regarded with interest: this adds further motivation for the present study.
However, at ultra-high energies the detection principle has to change, since we are dealing with distances that exceed the size of any conceivable detector: 
$$\ell_\tau =
c  \gamma_\tau \tau^0_\tau \approx 10\mbox{ km} \times\frac{E_{\tau}}{\mbox{0.2 EeV}} $$ 
Let us consider tau neutrinos of similar ultra-high energies, that interact inside the Earth and decay outside of it.
These could be observed by a detector devoted to monitor the circular region {\em below} its own horizon~\cite{fargion,iori,Fargion:2017rkg}; the recent efforts to achieve this goal using a satellite are documented in \cite{poe}.
Note that for energies  $\gtrsim 0.1$ EeV the tau particles interact significantly\footnote{\label{fooo}The range in water is $x\sim x_*  \log[ (1+E_\tau^{\mbox{\tiny in}} / \varepsilon)/(1+E_\tau^{\mbox{\tiny fin}} / \varepsilon)]$ with 
$x_*\sim 50$ km and 
$\varepsilon\sim 10$ TeV, much more than the decay length $\ell_\tau$ till 0.1 EeV. 
In fact, we have roughly 
$dE_\tau/dx=-(\alpha+\beta E_\tau)$, where $\alpha$ is almost the same as for the muon while $x_*=1/\beta$ 
(mainly due to pair production and to photonuclear interaction)
and $\varepsilon=\alpha/\beta$ scale roughly as $m_\tau/m_\mu$.}    
with the matter in which they propagate, see~\cite{luu} for a recent study.

\section{Results} \label{sec:IV}
\subsection{Parent function}
\begin{figure}[t]
\centering
\includegraphics[width=0.6\textwidth]{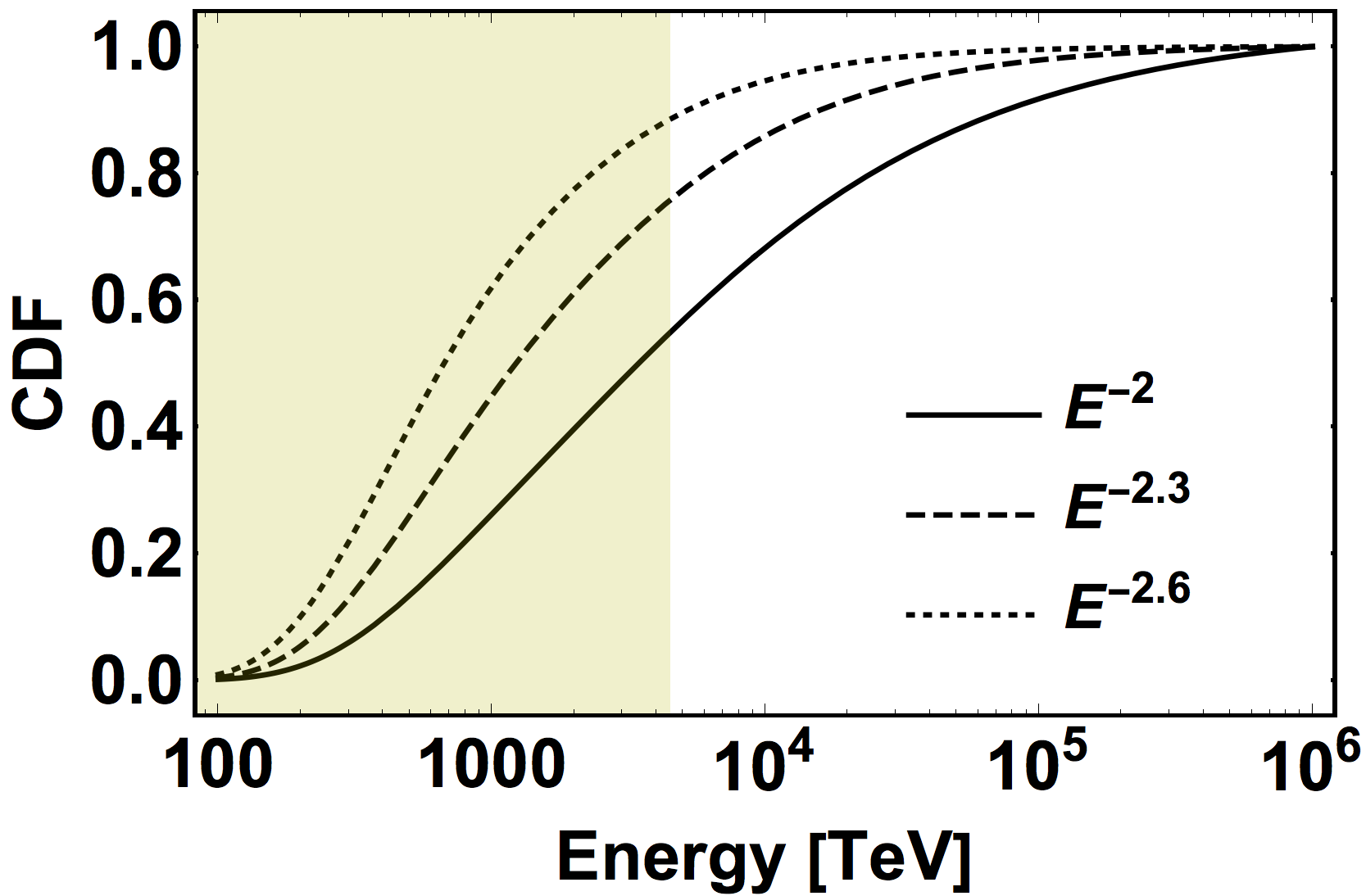}
\caption{Parent function of double cascade events (as defined in Eq.~\eqref{eq:par}), assuming power law spectra $E^{-\alpha}$, with $\alpha=2,2.3,2.6$, without any energy cutoff. The shaded region is the one currently explored by IceCube.}
\label{fig:par}
\end{figure}

The flux of high energy neutrinos, that is relevant for double pulse and double bang events, is the flux above few hundreds of TeV. 
This can be clearly seen looking at Fig.~\ref{fig:par}, where the parent functions of double cascade events are represented for different spectral indices. 
The parent function is defined as follows:
\begin{equation}
 P(E_\nu,\alpha)= \frac{\displaystyle\int_0^{E_\nu} E^{-\alpha} \, A_{\text{eff}}(E_\nu) dE_\nu}{\displaystyle\int_0^\infty E^{-\alpha} \, A_{\text{eff}}(E_\nu) dE_\nu}
 \label{eq:par}
\end{equation}
The plots in Fig.~\ref{fig:par} clearly show that, whatever is the spectral index of the neutrino spectrum, double cascades are mostly generated by neutrinos with energy above few hundreds of TeV. 
Therefore, when we generally discuss \lq\lq double cascade\rq\rq events, we can say that
\begin{quote}
the low energy part of the cosmic neutrino spectrum (below 200 TeV) is 
irrelevant for the prediction of the $\tau$ event rate.
\end{quote}
The previous consideration permits us to use directly the throughgoing muon flux \cite{icemuon}, avoiding all the discussion related to the tension between this spectrum and the HESE spectrum \cite{icecombined}, that shows a different behavior below 200 TeV.

Moreover, it is important to stress that tau neutrinos \textit{must} be observed \andreamod{assuming standard oscillations}, and that even the presence of a possible energy cutoff is not expected to modify this conclusion strongly. 
This statement is based on the fact that one 4.5 PeV track event has already been observed; this most likely means that there is no energy cutoff below this energy. 
In Fig.~\ref{fig:par} we notice that a considerable fraction of double pulse events is produced by neutrinos below 4.5 PeV; namely, between 60\% and 90\% going from an hard spectrum to a soft spectrum.  
Moreover, a 4.5 PeV track requires a more energetic neutrino to be produced, around 10 PeV or above (with large uncertainties, see \cite{pallwinter} where the energy reconstruction is widely discussed).

\subsection{Expected number of events in the pion decay scenario}
\
\label{sec:res1}
In this section we use the measured flux of throughgoing muons \cite{icemuon} to evaluate the expected number of double cascades in the three detectors as a function of the spectral index. 
We perform the calculation in two different ways: 
\begin{enumerate}
\item in this section we calculate the expectation for a particular case, i.e.\ the pion decay, in which the approximation $\phi_\tau = \phi_\mu$ is valid. 
Moreover, we show the expectations as a function of the spectral index; 
\item in the next section we show the general result, taking into account the uncertainties given by the normalization, the spectral index and the production mechanism.
\end{enumerate}

The normalization at 100 TeV, that we denote by $F(\alpha)$, changes with the spectral index, as reported in Fig.~6 of \cite{icemuon}. 
Namely, the normalization assumes the values 0.65, 0.8, 1, 1.25, 1.5, 1.75 (in the usual units of \SI{e-18}{\per\giga\electronvolt\per\square\centi\meter\per\second\per\steradian}) for spectral indices $2,2.1,\ldots,2.5$ respectively.

\begin{figure}[t]
\centering
\includegraphics[width=0.6\textwidth]{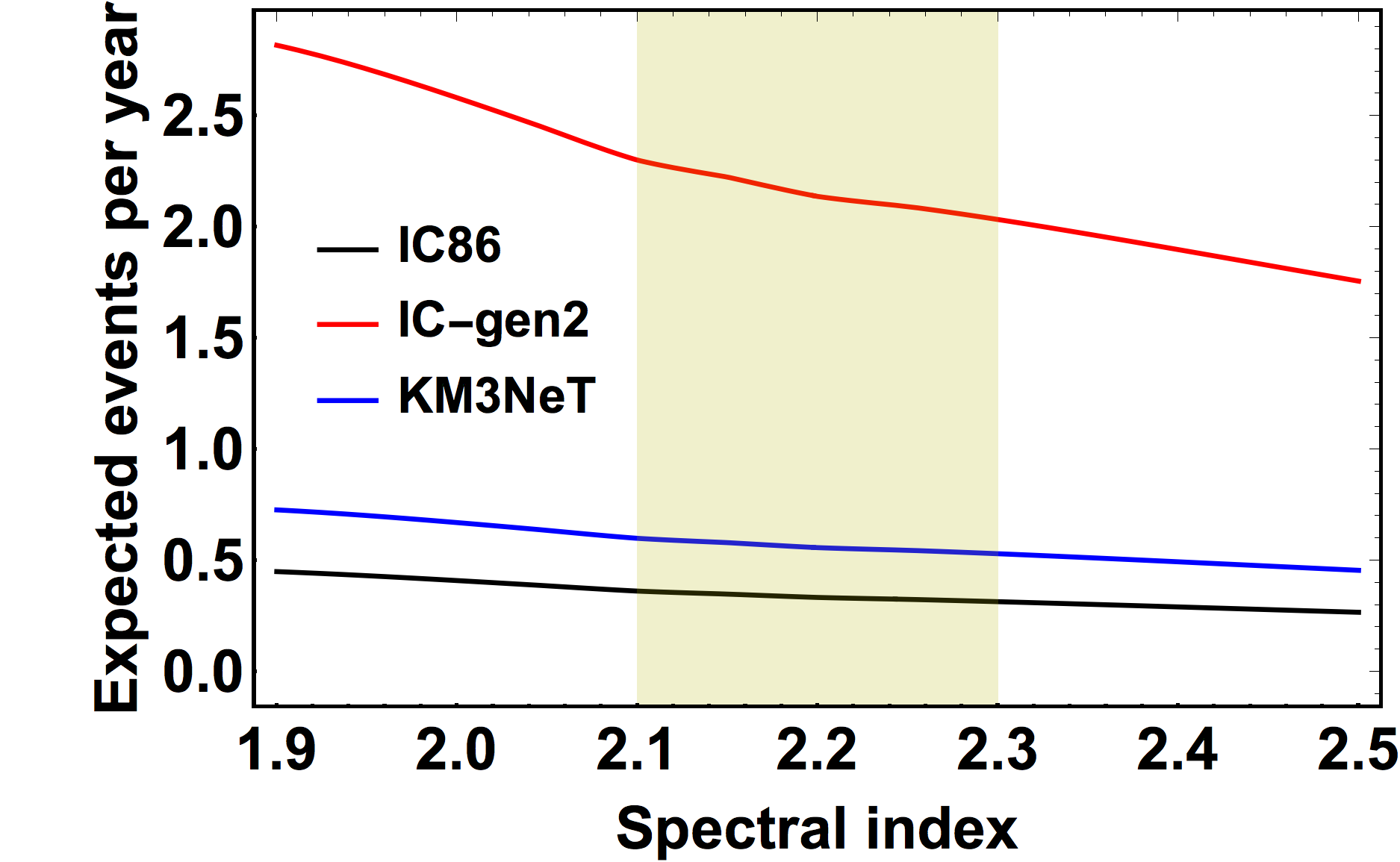}
\caption{Expected number of double cascades in the three detectors considering different spectral indices. \andreamod{The shaded region brackets spectral indices favored by the analysis} of the throughgoing muon events within 1$\sigma$. \andreamod{The shallow dip seen in the curves around $\alpha=2.1$ is due to the fact that the number of events does not decrease linearly for increasing spectral index. This behavior is related to the low energy threshold of the effective areas of double bang and double pulse; this effect can be also appreciated by looking at Fig.~\ref{fig:par}. }}
\label{fig:spec}
\end{figure}

We can use this information to evaluate the expected number of events per year as a function of the spectral index, using the usual formula:
$$
N_{\text{2casc}}^i(\alpha)= 4 \pi \text{T} \int_0^\infty  F(\alpha) \,E_\nu^{-\alpha} \, A_{\text{eff},\text{2casc}}^{i}(E_\nu)  \, dE_\nu
$$
where T is the exposure time and $A_{\text{eff},\text{2casc}}^{i}(E_\nu)$ is the effective area of the $i$-th experiment.
The results are reported in Fig.~\ref{fig:spec}, where the $3\sigma$ interval of the spectral index is shown, i.e.\ $1.9<\alpha<2.5$. 
In the same figure also the $1\sigma$ band (in yellow) of the spectral index is shown. 
Within the $1\sigma$ region the expected rate is roughly 0.35, 0.55 and 2.15 events per year in IceCube, KM3NeT and IceCube-gen2 respectively. 
The expectation changes of few \% within the 1$\sigma$ band and of about 40\%-50\% in the extreme intervals of the 3$\sigma$ band.

\subsection{Expected number of events: general case}
In this section we present the expectations, in the three different detectors, in the most general way. 
More specifically:
\begin{itemize}
\item we do not assume any spectral index, as we just use its probability distribution function $P(\alpha$), that is a Gaussian function, being $\alpha=2.19 \pm 0.1$;
\item we do not assume any specific production mechanism, using the distribution $R_{\tau\mu}$ defined in Sec.~\ref{sec:II}.
\end{itemize}
Under these hypotheses the expected number of events is given by the following formula:
\begin{equation}
\braket{N_{\text{2casc}}^i}= 4 \pi \text{T} \int_0^\infty dE_\nu \int_{0}^\infty d\alpha \int_0^\infty dr\, 
r  F(\alpha) \, E_\nu^{-\alpha} \,  A_{\text{eff},\text{2casc}}^{i}(E_\nu) \, P(\alpha)\,  \frac{d \rho}{d r}(r)
\end{equation}
where $\phi_\tau(E_\nu)= r F(\alpha) E_\nu^{-\alpha}$ is the flux of tau neutrino, and we indicate $\phi_\tau/\phi_\mu$ as $r$, rather than $R_{\tau\mu}$ as in Eq.~\eqref{eq:rtaumu}, to shorten the notation.
The function $d\rho/dr$ is the normalized distribution of $R_{\tau\mu}$, shown in the rightmost in Fig.~\ref{fig:taufrac}.
\andreamod{The meaning of the integral in $\alpha$ and $r$ is that, in absence of more precise information on these parameter, we take into account the current uncertainty on the slope of the neutrino distribution and on the flavor ratio, weighting the parameters in the most unbiased manner\footnote{\andreamod{With improved experimental and/or theoretical knowledge, this information should be updated.}}.}
$A_{\text{eff},\text{2casc}}^{i}(E_\nu)$ is the effective area for double cascades (i.e.~double pulses + double bangs) for the detector $i$-th, T is the exposure time, $P(\alpha)$ is the PDF of the spectral index and $R_{\tau\mu}(r)$ is the distribution of the ratio between $\phi_\tau/\phi_\mu$ in the case of a generic production mechanism at the source, see the rightmost panel of Fig.~\ref{fig:taufrac}.

\begin{table}[t]
\centering
\caption{In the columns from 2\carlo{nd} to 4th, the expected yearly rates of tau neutrino events, obtained integrating over the spectral index distribution and over the production mechanisms. 
The uncertainties associated to these expectation are 30\%. 
In the column from 5th to 7th the number of years required to observe at least one double cascade with a certain probability, namely 90\%, 99\% or 99.9999\% (5$\sigma$).
In this calculation we consider that the background is equal to the 40\% of the signal, \andreamod{as plausibly expected from the information contained in \cite{icrc}.}}
\label{tab:finres}
\begin{tabular}{lccccrr}
\midrule
Experiment & $N_{\text{2bang}}$ & $N_{\text{2p}}$ & $N_{\text{2casc}}$ & $T_{\text{year}}^{\text{P}>90\%}$ &$T_{\text{year}}^{\text{P}>99\%}$ & $T_{\text{year}}^{\text{P}>5\sigma}$  \\
\midrule
IC86 & 0.07 & 0.25 & 0.32 & 5.1 & 10.1 &31.7\\
IC-gen2 & 0.29 & 1.78 & 2.07 & 0.8 &1.6 &  5.0\\
KM3NeT & 0.10 & 0.44 & 0.54 &3.1 & 6.1 & 19.1 \\
\midrule
\end{tabular}
\end{table}

The expected yearly rates of double cascades for IceCube, IceCube-gen2 and KM3NeT are, respectively, 0.32, 2.07 and 0.54. 
Other details are reported in Tab.~\ref{tab:finres}. 

The ratio between double pulse and double bang events is 4:1 in IceCube and KM3NeT, whereas it becomes 6:1 in IceCube-gen2, as the larger distance between the strings disfavors the double bang detection.

The uncertainties produced by the spectral index and by the production mechanism can be computed using the following formula:
\begin{equation}
\braket{\Delta N_{\text{2casc}}^i} = \left(\int_{0}^\infty d\alpha \int_0^\infty dr \,r \left[ N_{\text{2casc}}^i(\alpha) - \braket{N_{\text{2casc}}^i}\right]^2 \, P(\alpha) \, \frac{d\rho}{dr}(r)\right)^{1/2}
\end{equation}
where $N_{\text{ev}}^i(\alpha)$ are the expected number of double cascade events for an $F(\alpha) E^{-\alpha}$ spectrum, evaluated in Sec.~\ref{sec:res1}. 
The uncertainties due to the spectral index and the production mechanism amount to $\sim 10\%$, therefore the global uncertainty is dominated by that on the spectrum normalization ($\sim 25\%$ \cite{icemuon,icrc}), and it is equal to 30\%, summing the uncertainties in quadrature.

\begin{figure*}[t]
\centering
\includegraphics[scale=0.6]{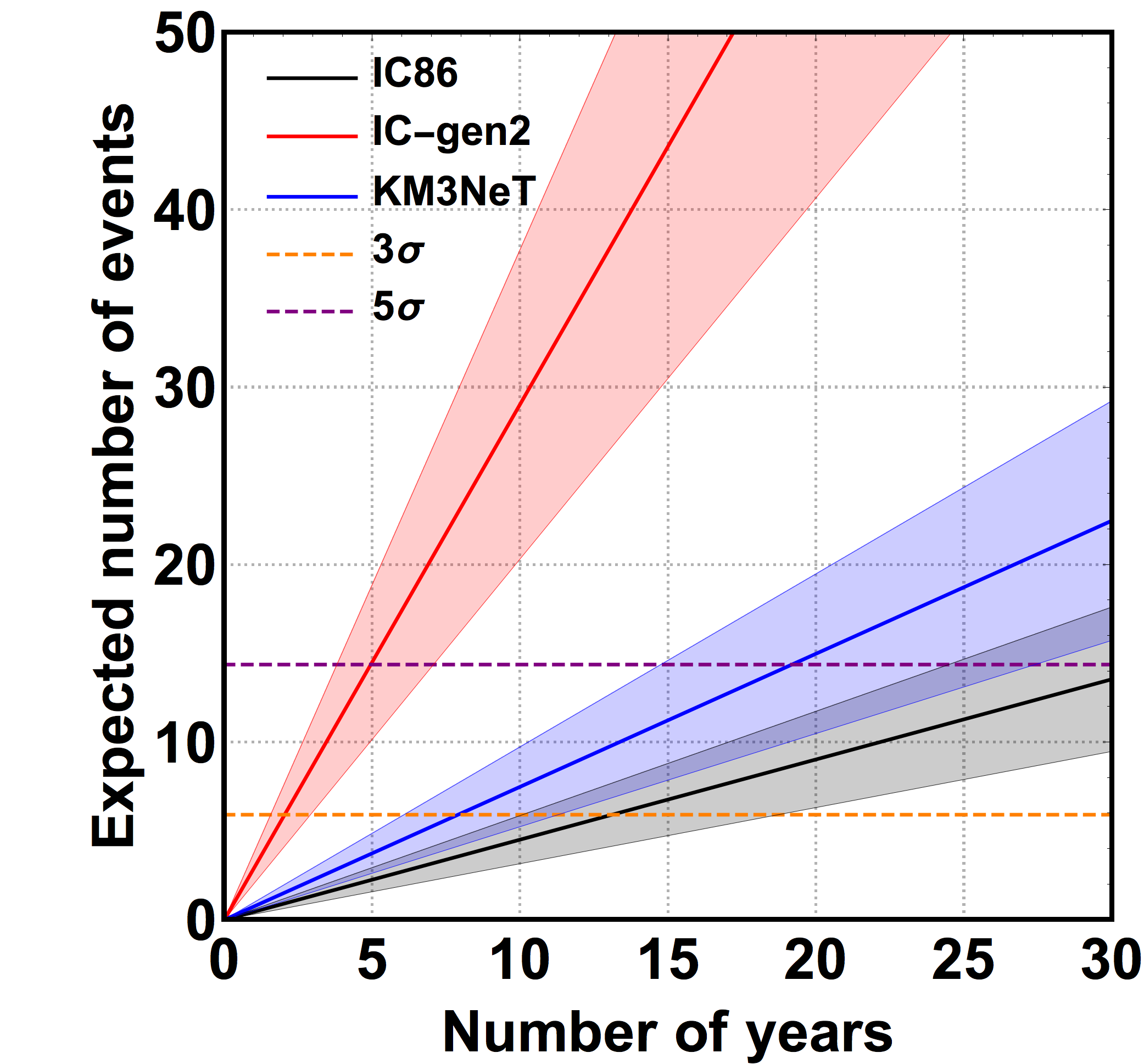}
\caption{The expected number of double cascade events in the three detectors, assuming \andreamod{ that the background is equal to the 40\% of the signal. The colored bands represent the global uncertainties (spectral index + normalization + production mechanism).}}
\label{fig:res}
\end{figure*}

\subsubsection{Expected number of background events}
At this point a clarification is necessary. 
The flux of $\nu_\tau$ produced in atmosphere is very low compared to the astrophysical flux of $\nu_\tau$, because atmospheric tau neutrinos are produced in the decay of the \carlo{rare} meson $D_{s}$. 
Following \cite{Bulmahn:2010pg}, the flux of atmospheric tau neutrinos is approximately equal to:
$$
E^2 \phi_{\nu_\tau}^{\text{atmo}} \simeq 2 \times 10^{-10} \left(\frac{E}{100 \mbox{ TeV}} \right)^{-0.7} \si{\giga\electronvolt\per\square\centi\meter\per\second\per\steradian} 
$$
Using the usual effective area for double cascade events, we find that the expected rate of double cascades produced by atmospheric $\nu_\tau$ is 1/200 of the rate due to astrophysical neutrinos. 
Therefore this source of background is totally negligible.

On the contrary an important source of background is represented by the misidentified events, classified as double cascades. 
At present, it is not trivial to evaluate this aspect, \andreamod{since the number of misidentified double cascades is not linearly related to other measured atmospheric backgrounds, but it also depends on experimental details connected to the technology used to detect high energy neutrinos. The best that we can do is to use the available informations presented in
\cite{icrc}, where the misidentified double cascades are expected to be not negligible and roughly equal to the $\sim 40\%$ of the number of double cascades produced by astrophysical neutrinos.}

In Fig.~\ref{fig:res} we show the expected number of events as a function of the number of years, showing also the associated global uncertainties with the colored bands. 
\andreamod{These predictions are obtained assuming that the background is 40\% of the signal, as supported by the informations contained in \cite{icrc}.}
Let us discuss how many years are required to observe at least 1 event with a certain probability: 
\begin{itemize}
\item the present IceCube should be close to identify a tau neutrino with a probability of 90\%, since an exposure of 5.1 years is required and 6 years of HESE have been already collected.
On the other hand, if IceCube does not observe any $\nu_\tau$ in the next years, this would be still compatible within 5$\sigma$ with the theory, until $\sim 30$ years of exposure;
\item the situation is totally different in IceCube-gen2, where a $\nu_\tau$ should be identified in $\sim 1$ year with a probability of 99\%.
The non observation of any double cascade would contradict the theory at the level of 5$\sigma$, after only 5 years;
\item the future KM3NeT should have slightly better performance than IceCube, due to the fact that the strings are separated by 90 meters and water is $\sim 10\%$ denser than ice. 
This experiment should observe a tau neutrino in about 3 years with a probability of 90\%. 
On the other hand, as for the present IceCube, the non observation of double cascades would represent an issue at 5$\sigma$ only after an exposure of $\sim$ 20 years.
\end{itemize}

In conclusion, for both IceCube and KM3NeT some tens of years worth of data have to be collected, before being in contradiction with the theory if no tau neutrinos will be observed.
For these reasons the role of IceCube-gen2 is crucial for the observation of this kind of events, since it is expected to observe 2 double cascade events per year in the most general scenario. 
All these results are summarized in Tab.~\ref{tab:finres}.

The non observation of tau neutrinos would have dramatic consequences for the neutrino physics, such as:
\begin{itemize}
\item the fraction of $\nu_\tau$ is much smaller than what expected.
Therefore $\phi_\tau$ is not connected to the flux of $\phi_\mu$, which would mean that neutrino oscillations are violated. \andreamod{This scenario would imply evidence of new physics.}
\item tau neutrinos are not observed because neutrino telescopes are observing mostly atmospheric background, in which tau neutrinos are not present. 
This would mean that cosmic neutrinos have not been observed.
\end{itemize}
For these reasons the direct observation of tau neutrinos is a crucial point for neutrino physics, and their eventual non observation cannot be overlooked in the next years.

\subsection{Discrimination between signal and misidentified double cascades}  \label{sec:IV-4}

In the previous sections we computed the expected number of double cascades produced by astrophysical neutrinos. 
We have estimated and taken into account the background events resulting from atmospheric $\nu_\tau$, that, as discussed, are very few, fewer than 1\% of the signal events. 
We also discussed that the most important source of background is related to the misidentified events.  
The misidentified events, classified as double cascades, become relevant when we want to know how many years are required to extract the astrophysical signal from the sample of observed events. 
This will become the most important task, once a sufficient sample of double cascade events will be detected. 

As in the previous section and as quoted in \cite{icrc}, the background rate, which exists for sure, corresponds to about 40\% of the signal one, i.e.~$N_{b}\approx 0.13\;\text{yr}^{-1}$ for the present IceCube.
If we use this value, it is evident that it is not possible yet to discriminate the signal and the background in IceCube and KM3NeT, since, as indicated in Tab.~\ref{tab:finres}, the rate of data collection is expected to be very low. 
On the other hand, the separation between the double cascades produced by astrophysical neutrinos and those due to misidentification is achievable in IceCube-gen2. 

\begin{figure}[t]
\centering
\includegraphics[width=0.6\textwidth]{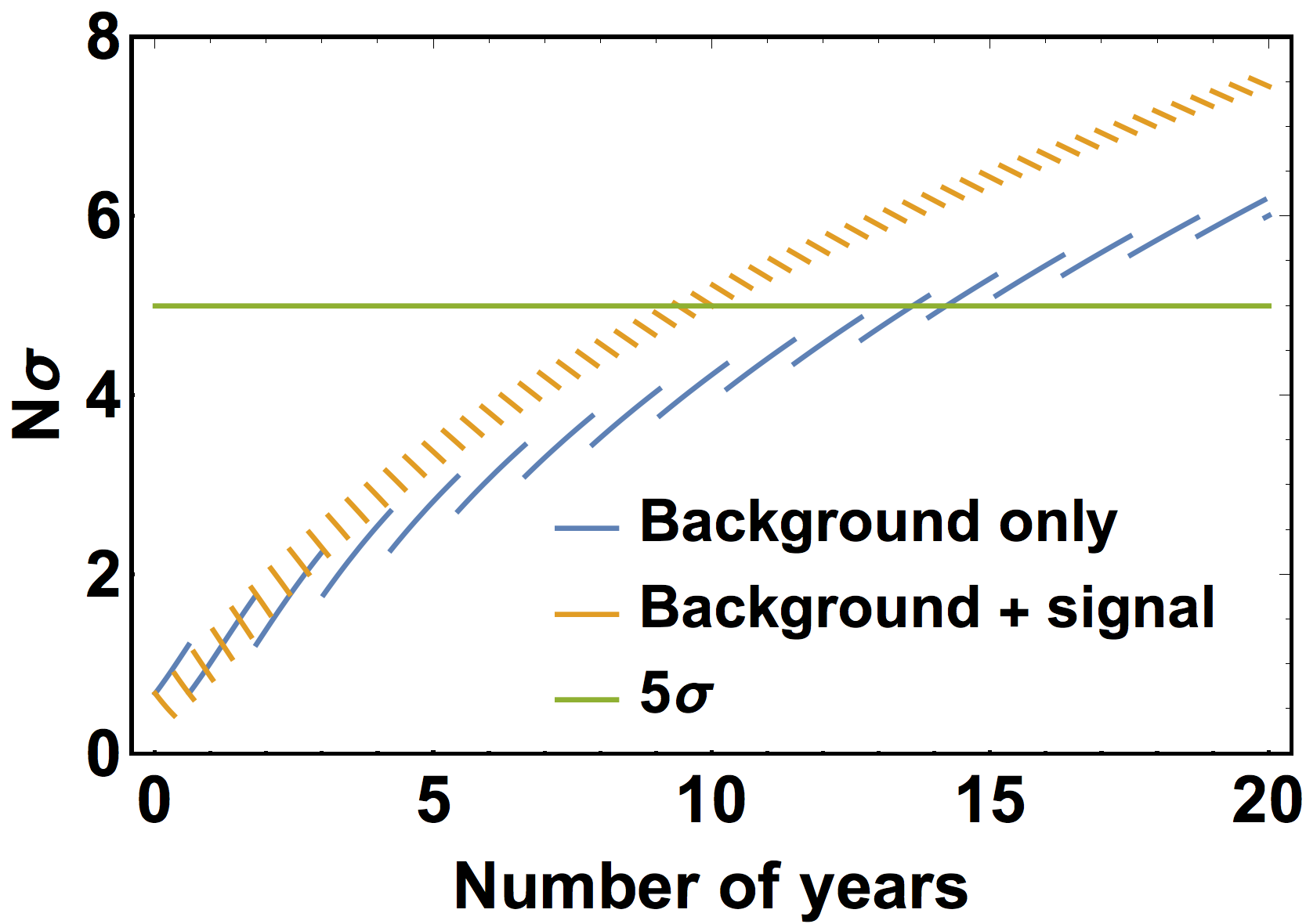}
\caption{In this plot we represent the function TS, defined in Eq.\ \ref{eq:TS}, for the following cases: \textit{i)} background only (blue line), \textit{ii)} background + signal (orange line). \andreamod{The broken curves follow formally from the function `integer part' in Eq.~(\ref{porchetton}), that describes the fact that, as the time goes by, new individual events are expected to accumulate according to the specified hypothesis.}
}
\label{plotstatvis}
\end{figure}

In order to test whether the (future) data are consistent with the background, which we call hypothesis $H_0$, or instead
indicate the presence of a signal along with the background, 
which we call hypothesis $H_1$, 
it is useful to define the following test statistic:
\begin{equation}
\mbox{TS}=\frac{\mathcal{P}(n|H_0)}{\mathcal{P}(n|H_0)+\mathcal{P}(n|H_1)}
\label{eq:TS}
\end{equation}
where $\mathcal{P}(n|H_i)$ is the conditional probability of observing $n$ events assuming the hypothesis $H_i$.
For us the yearly rate of signal events is $0.32\;\text{yr}^{-1}$, and the background rate is $40\%$ of that, i.e.~$0.13\;\text{yr}^{-1}$: the expected number of events after an exposure time $t$ is thus $\mu_0=0.13\,t/\si{yr}$ in the case of background only, and $\mu_1=0.32\,t/\si{yr}$ in the case of signal and background. 
We assume Poissonian distributions for $\mathcal{P}(n|H_i)$, i.e.:
$$\mathcal{P}(n|H_i)=\frac{\mu_i^n}{n!} e^{-\mu_i} $$
At present ($ t=5.7$ yr) 
no events are seen ($n=0$) while we expect $5.7 \mbox{year} \times 0.45 \frac{\mbox{events}}{\mbox{year}} =2.6$ events\footnote{Note that this prediction is based on the throughgoing muon spectrum (see also Tab.\ref{default}) and for this reason slightly differs from the prediction of \cite{icrc} where an $E^{-2.5}$ spectrum is considered.} and thus,
$$
\left. \text{TS}\right|_\text{\tiny today}=\left( 1+e^{-0.45\,t/\si{yr}} \right)^{-1}= 93\%
$$ 
This is not worrisome at all, as this value for the TS corresponds to a 1.8$\sigma$-level evidence of background only, but we think the community should be aware of how fast this could change with IceCube-gen2. 
In order to estimate the exposure needed to accept the hypothesis $H_0$ or to reject the hypothesis $H_1$ at a certain confidence level,  it is convenient to assume two cases: \begin{enumerate} 
\item over time data accumulate  
as expected in the case of background only;
\item also the signal is present.
\end{enumerate}
These cases are 
described by the assumption:
\begin{equation} \label{porchetton}
n(t)= \begin{dcases}
[ N_{{b}}\, t + 1/2 ] & \text{background only} \\
[(N_{{b}}+N_{{s}})\,t+ 1/2 ] & \text{also signal} 
\end{dcases}
\end{equation}
where $[x]$ denotes the integer part of $x$. 
The result for the test statistics is shown in Fig.~\ref{plotstatvis}, which concerns IceCube-gen2 \cite{ice2}. 
From this figure we see that if the signal is actually present and it is correctly described, it will take 10 years to confirm it at 5$\sigma$, 
while it will take 15 years to establish the most pessimistic scenario, when  the signal is absent.\footnote{A different and simpler take on this matter is the following. 
We can obtain a similar result by comparing the expected number of background events with an under-fluctuation of the expectations in the case when the signal is also present. 
If $N_b$ and $N_s$ are the background and signal rates respectively, we obtain that the condition
$$
N_\text{b} t < (N_\text{s}+N_\text{b}) t -5 \sqrt{(N_\text{s}+N_\text{b})  t }
$$ 
which corresponds roughly to the 5$\sigma$ criterion, is satisfied after 17 years.}
The meaning of this result is the following:
\begin{quote}
once double cascade events will be detected, in order to claim with great confidence that cosmic tau neutrinos have been observed, an important exposure will be required.
\end{quote}
On the other hand, as discussed in the previous sections, the non observations of any double bang in IceCube-gen2 could become an issue since the occurrence of double cascade events are expected with a probability of 99\% already after 2 years.

Let us remark that the results presented in this section have the character of estimation and might improve by subsequent experimental work and systematic analysis: a significant reduction of the expected misidentified events will be important to decrease the required exposure.

\section{Summary}
\label{sec:V}
In this work we discussed the importance to observe tau neutrinos, both in the present and, mostly, in the future neutrino telescopes. 
Tau neutrinos are expected to contribute from 20\% to 40\% to the total astrophysical neutrino flux, while they are not significantly produced as atmospheric secondary particles. 
We remarked that the knowledge of cosmic muon neutrinos allows us to derive a reliable prediction on the associated flux of tau neutrinos, simply owing to the known three flavor neutrino flavor oscillations.
Unfortunately, it is quite hard to identify tau neutrinos, as they have distinguishable signatures in neutrino telescopes only at very high energy, above few hundreds of TeV. 

In Sec.~\ref{sec:I} and Sec.~\ref{sec:II} we have proposed a brief review of the status of the art of neutrino oscillations.
We have argued that there are excellent reasons to believe that they occur once high-energy neutrinos travel over cosmic scales while  there are no strong reasons to assume the existence of other phenomena; moreover, we have evaluated accurately the effect of neutrino oscillations. 
In Sec.~\ref{sec:III} we have proposed a general way to to parameterize the effective areas for double pulses and double bangs for a generic detector. 
Then we adapted our general parametrization to the neutrino telescopes that already exist and we estimated the effective areas of future neutrino telescopes. 

The effective areas of double bang events are not available in the literature. We have described how to obtain an approximate description from theoretical considerations, but we do not consider these estimations as reliable as the effective areas of double pulse events, that have been tested with those calculated by the experimental collaborations instead. On the other hand, the uncertainties on the former effective areas do not have an important impact on the total number of expected tau-neutrino events since, as discussed in Sec.~\ref{sec:IV}, the double pulses are at least 4 times more than double bangs. 

Using these effective areas we have computed in Sec.~\ref{sec:IV} the expected yearly rates of distinguishable tau neutrino events. 

Based on conventional physics and IceCube measurements, we conclude that the present IceCube detector is close to observe the first double cascade with a probability of 90\%.
On the other hand, the non observation of tau neutrinos will be in tension with the theory at 5$\sigma$ C.L.~only after $\sim 30$ years of exposure.
We have obtained a similar result for KM3NeT, where about 3 years are required to observe a double cascade with a 90\% C.L., but about 20 years of non observation are required to be in tension with the theory at 5$\sigma$. 
The situation is completely different for IceCube-gen 2, for which we expect the detection of a double cascade in $\sim 1$ year with a probability of 99\%. 
The non observation of double cascades would become problematic (at 5$\sigma$) after only $\sim 5$ years.
Let us remark that this predictions are very robust, since they only depend on the high energy part of neutrino spectrum, i.e.~above $\sim \SI{200}{TeV}$. 
This energy range has been measured by IceCube both with HESE and throughgoing muons, and the two measurements are in good agreement in this energy range. 
The tension between HESE and throughgoing muons is only present at low energies (below 100 TeV), but it 
does not affect the prediction for double cascades, as explained in the text. 
Moreover the uncertainties related to the production mechanism and to the spectrum (spectral index and normalization) are taken into account and they produce a total uncertainty of 30\% on the expectations.

The direct observation of tau neutrinos is a crucial issue 
for high energy neutrino astronomy and it should be regarded as a priority of the new generation of detectors. 
Indeed, we would like to conclude stressing that the {\em non-observation} of cosmic tau neutrino events in the next generation of neutrino telescopes would have dramatic consequences for the neutrino physics, possibly implying that: \\\textit{i)} neutrino oscillations are violated; \\ \textit{ii)} cosmic neutrinos have not been observed;\\ \textit{iii)} there is new 
and very unexpected physics to be explored.

\subsection*{Acknowledgments} We thank  R.~Aloisio, V.~Berezinsky and A.~Studenikhin for useful discussions. The work of A.~P.~has been supported with funds of  the European Research Council (ERC) under the European Union's Horizon 2020 research and innovation programme (Grant Number 646623).

\andreamod{
\subsection*{Note added (July 13, 2018)} 
After this study was issued (April 13), IceCube collaboration announced a couple of candidate tau events at the {\em Neutrino 2018} meeting \cite{ign} (June 6). Then a correlation between one high-energy neutrino event and one flaring blazar was reported  \cite{july12} (July 12). 
These are excellent news for high-energy neutrino astronomy; furthermore, they add credibility to a straightforward interpretation of IceCube findings, as the one adopted in the present work.}

\end{document}